\documentclass[11pt]{article}

\usepackage{palatino}
\usepackage{mathpazo}
\usepackage{amsmath}
\usepackage{amsthm}
\usepackage{stmaryrd}

\newtheorem{theorem}{Theorem}
\newtheorem{lemma}[theorem]{Lemma}

\newtheorem{cor}[theorem]{Corollary}
\theoremstyle{definition}
\newtheorem{definition}[theorem]{Definition}

\newtheorem{remark}[theorem]{Remark}

\newtheorem{problem}[theorem]{Problem}

\newcommand{\bra}[1]{\langle #1|}
\newcommand{\ket}[1]{|#1\rangle}
\newcommand{\tinyspace}{\mspace{1mu}}
\newcommand{\op}[1]{\operatorname{#1}}
\newcommand{\norm}[1]{\left\lVert\tinyspace#1\tinyspace\right\rVert}
\newcommand{\abs}[1]{\left\lvert\tinyspace #1 \tinyspace\right\rvert}
\newcommand{\dnorm}[1]{\norm{#1}_{\diamond}}
\newcommand{\cbnorm}[1]{\norm{#1}_{\mathrm{cb}}}
\newcommand{\tnorm}[1]{\norm{#1}_1}
\newcommand{\defeq}{\stackrel{\smash{\text{\tiny def}}}{=}}
\newcommand{\tr}{\operatorname{Tr}}
\renewcommand{\t}{{\scriptscriptstyle\mathsf{T}}}
\def\({\left(}
\def\){\right)}
\def\I{\mathbb{1}}
\newcommand{\fid}{\operatorname{F}}
\newcommand{\lin}[1]{\setft{L}\left(#1\right)}
\newcommand{\setft}[1]{\mathrm{#1}}
\newcommand{\density}[1]{\setft{D}\left(#1\right)}
\newcommand{\trans}[1]{\setft{T}\left(#1\right)}
\newcommand{\herm}[1]{\setft{Herm}\left(#1\right)}
\newcommand{\pos}[1]{\setft{Pos}\left(#1\right)}
\newcommand{\sphere}[1]{\setft{S}\left(#1\right)}
\newcommand{\opset}[3]{\setft{#1}_{#2}\left(#3\right)}

\def\complex{\mathbb{C}}
\def\real{\mathbb{R}}

\def\X{\mathcal{X}}
\def\Y{\mathcal{Y}}
\def\Z{\mathcal{Z}}
\def\W{\mathcal{W}}
\def\A{\mathcal{A}}

\def\Q{\mathcal{Q}}
\def\U{\mathcal{U}}
\def\V{\mathcal{V}}

\begin{document}

  \title{\bf 
    Distinguishing quantum operations\\ 
    having few Kraus operators}
  
  \author{
    John Watrous\\[1mm]
    {\it\small 
      Institute for Quantum Computing and School of Computer Science}\\
    {\it \small University of Waterloo, Waterloo, Ontario, Canada.}
  }
  
  \date{\normalsize April 18, 2008}

  \maketitle
  
  \begin{abstract}
    Entanglement is sometimes helpful in distinguishing between
    quantum operations, as differences between quantum operations can
    become magnified when their inputs are entangled with auxiliary
    systems.
    Bounds on the dimension of the auxiliary system needed to
    optimally distinguish quantum operations are known in several
    situations.
    For instance, the dimension of the auxiliary space never needs to
    exceed the dimension of the input space \cite{Smith83,Kitaev97} of
    the operations for optimal distinguishability, while no auxiliary
    system whatsoever is needed to optimally distinguish unitary
    operations \cite{AharonovK+98,ChildsPR00}.
    Another bound, which follows from work of R.~Timoney
    \cite{Timoney03}, is that optimal distinguishability is always
    possible when the dimension of the auxiliary system is twice the
    number of operators needed to express the difference between the
    quantum operations in Kraus form.
    This paper provides an alternate proof of this fact that is based
    on concepts and tools that are familiar to quantum information
    theorists.
  \end{abstract}

  \maketitle
  
\section{Introduction}

The notion of entanglement is pervasive in the theory of quantum
information, often playing a critically important and yet sometimes subtle
role in different settings.
One such setting concerns the distinguishability of quantum
operations, which has been considered in various forms by several
authors
\cite{%
  Acin01,
  AharonovK+98,
  ChildsPR00,
  D'ArianoPP01,
  GilchristLN05,
  Kitaev97,
  KretschmannSW06,
  RosgenW05,
  Sacchi05,
  Sacchi05b}.

Consider a situation in which two quantum operations $\Phi_0$ and
$\Phi_1$ are fixed.
A single evaluation of one of the two operations is given, and the
goal is to determine which of the two operations it is.
This type of problem will be considered in greater generality
momentarily, but for the moment assume that $\Phi_0$ and $\Phi_1$ are
single-qubit operations. 
Also assume that a bit $a\in\{0,1\}$, chosen uniformly at random,
determines which of the two operations is given, so that it is
meaningful to consider the optimal probability with which the given
operation is correctly identified.

A natural approach to an instance of this problem is to optimally
choose a single-qubit input state $\rho$ so that the output states
$\Phi_0(\rho)$ and $\Phi_1(\rho)$ are as far apart as possible (with
respect to the trace norm, for instance).
Then, some optimal measurement can be applied to the output state
$\Phi_a(\rho)$ to obtain information about the bit $a$. 

This, however, is not the most general approach, and is not always optimal.
More generally, one may prepare a possibly {\it entangled} state
between the input to the operation and some auxiliary system, and
then apply the operation $\Phi_a$ to the input system.
A multiple-qubit measurement may then be applied to the output and
auxiliary systems together to obtain information about~$a$.
Indeed this more general approach can give an improvement in the
probability of correctly identifying the bit $a$ in some cases.

For example, consider an instance of the above problem in which $\Phi_0$
is the identity operation, while $\Phi_1$ corresponds to the
application of a randomly chosen non-identity Pauli operator:
\[
\Phi_1(\rho) = \frac{1}{3} \sigma_x \rho \sigma_x 
+ \frac{1}{3} \sigma_y \rho \sigma_y 
+ \frac{1}{3} \sigma_z \rho \sigma_z.
\]
These two quantum operations can be distinguished without error using
an entangled input state as follows:
any one of the four Bell states is chosen, $\Phi_a$ is applied to one
of a pair of qubits in this state, and the two qubits are
measured with respect to the Bell basis.
In case $a = 0$, the result of the measurement obviously agrees with
the initially chosen Bell state, while in case $a = 1$, the result of
the measurement will correspond to one of the three remaining Bell
states, never resulting in the initially chosen state.
In this way, the index $a$ can be identified without error, and so
$\Phi_0$ and $\Phi_1$ can be distinguished perfectly.
Perfect distinguishability of $\Phi_0$ and $\Phi_1$ is, however, not
possible with a strategy that does not entangle the input to the
operations with an auxiliary system: the optimal probability of
correctly guessing $a$ with such a strategy can be shown to be~$5/6$.

A related example is discussed later in Section~\ref{sec:example} that
illustrates that a striking gap can exist between the entangled and
non-entangled approaches to this problem.
(It is nearly the same as an example that was discussed in
\cite{KretschmannSW06}.)
In particular, quantum operations acting on large systems can
sometimes be distinguished perfectly using entanglement with an
auxiliary system, and yet act nearly identically on inputs not
entangled with an auxiliary system.
A similar phenomenon arises in the context of approximate
randomization of quantum states \cite{HaydenLSW04}.

It is, however, not always the case that entanglement with an
auxiliary system helps in this problem.
While it is easy to construct trivial examples of this sort, there is
an interesting general class of examples known:
if $\Phi_0$ and $\Phi_1$ are arbitrary {\it unitary} operations, then
optimal distinguishability is possible without an auxiliary system
\cite{AharonovK+98,ChildsPR00}.
The same fact holds more generally when $\Phi_0$ and $\Phi_1$ are
given by $\Phi_0(X) = A X A^{\ast}$ and $\Phi_1(X) = B X B^{\ast}$ for
linear isometries $A$ and $B$.

In light of these examples, it is natural to ask how large an
auxiliary system is needed for optimal distinguishability between
various classes of quantum operations.
In general, it is known that optimal distinguishability never requires
an auxiliary system that is larger than the input space of the
operations \cite{Smith83,Kitaev97}, while the example to be discussed in
Section~\ref{sec:example} shows that the probability to distinguish
operations can sometimes shrink with even a small decrease in the size
of the auxiliary system from this upper bound.

This paper focuses on a lesser-known (and incomparable) bound: for
quantum operations $\Phi_0$ and $\Phi_1$, it is sufficient for optimal
distinguishability that the dimension of the auxiliary system is twice
the number of Kraus operators needed to express the difference between
$\Phi_0$ and $\Phi_1$, which is at most twice the total number of
Kraus operators needed to express $\Phi_0$ and $\Phi_1$.

Note that this bound is independent of the size of the systems the
quantum operations act upon, and may be viewed as a generalization of
the above-mentioned fact that unitary operations require no auxiliary
systems for optimal distinguishability.
That particular fact, however, is not quite recovered, for the bound
obtained only establishes that at most two auxiliary qubits are
required in this particular case rather than zero.
The bound is also clearly not interesting in the case where the
difference between the quantum operations to be distinguished requires
a number of Kraus operators that exceeds the dimension of the input
space of the operations. 
Nevertheless, the results hold generally for all quantum operations,
and may potentially be of use in understanding quantum operations with
few Kraus operators.
Recent work on quantum expanders
\cite{Ben-AroyaT-S07,Ben-AroyaST-S07,GrossE07,Harrow07,Hastings07,Hastings07b}
provides a setting where quantum operations with few Kraus operators
are of interest for some applications.

The above bound follows from a theorem of Timoney \cite{Timoney03},
whose proof is based on the notion of the Haagerup estimate on the
norm of complete boundedness for a class of super-operators on
$C^{\ast}$-algebras.
This paper provides a different proof based on notions that are
familiar in the theory of quantum information.
In particular, the well-known {\it fidelity} function plays a central
and simplifying role in the proof.
One of the technical parts of the proof, based on a theorem of
Barvinok \cite{Barvinok02}, may also be of independent use in quantum
information theory: every non-zero output of a
positive super-operator, ranging over all density operator inputs,
must have a low-rank preimage.

The remainder of the paper is organized as follows.
Section~\ref{sec:background} reviews background material needed for
the paper, including a discussion of super-operator representations
and distinguishability.
Section~\ref{sec:example} gives an example of quantum operations
that require a large auxiliary system to be distinguished optimally.
The actual bound discussed above on the size of the auxiliary space
needed for optimal distinguishability of quantum operations is proved
in Section~\ref{sec:main}. 

\section{Background} \label{sec:background}

\subsection{Basic linear algebra}

In this paper the term {\it complex Euclidean space} refers to any 
finite dimensional inner product space over the complex numbers
$\complex$, and we assume that every such space has a fixed orthonormal
{\it standard basis}.
For the remainder of this section, let $\X$ and $\Y$ be arbitrary
complex Euclidean spaces, and let $\{\ket{a}\,:\,a\in\Sigma\}$ 
denote the standard basis of $\X$, with $\Sigma$ being some arbitrary
finite, non-empty set.

The space of (linear) operators mapping $\X$ to $\Y$ is denoted
$\lin{\X,\Y}$, while $\lin{\X}$ is shorthand for $\lin{\X,\X}$.
The adjoint (or Hermitian transpose) of $A\in\lin{\X,\Y}$ is denoted
$A^{\ast}$, and the identity element of $\lin{\X}$ is denoted $\I_{\X}$.
If $\V$ is a subspace of $\X$, we let $\Pi_{\V}\in\lin{\X}$
denote the orthogonal projection onto $\V$.
We write $\herm{\X}$ to refer to the set of Hermitian operators on
$\X$, $\pos{\X}$ to refer to the set of positive semidefinite
operators on $\X$, and $\density{\X}$ to refer to the set of density
operators on $\X$.
The notation $A\geq 0$ also means that $A$ is positive semidefinite,
and more generally $A\geq B$ means that $A - B$ is positive semidefinite.

The {\it spectral norm} of an operator $A\in\lin{\X,\Y}$ is defined as
\[
\norm{A} = \max\{ \norm{A u}\,:\,u\in\sphere{\X}\}
\]
where $\sphere{\X} = \{u\in\X\,:\,\norm{u} = 1\}$
denotes the unit sphere in $\X$.
The {\it trace norm} of an operator $A\in\lin{\X,\Y}$ is defined as
\[
\tnorm{A} = \tr \sqrt{A^{\ast} A}.
\]
Equivalently, $\tnorm{A}$ is the sum of the singular values of $A$.

The {\it fidelity} between positive semidefinite operators
$P,Q\in\pos{\X}$ is defined as
\[
\fid(P,Q) = \tnorm{\sqrt{P}\sqrt{Q}} = \tr\sqrt{\sqrt{Q} P \sqrt{Q}}.
\]
This function has also been called the {\it tracial geometric mean}
in work of Timoney \cite{Timoney07} that is subsequent to the paper
\cite{Timoney03} that is most closely related to this one.

\subsection{Linear super-operators and representations}

A linear mapping of the form
$\Phi:\lin{\X} \rightarrow \lin{\Y}$
is a {\it super-operator}, and the space of all such mappings is
denoted $\trans{\X,\Y}$.
As expected, the notation $\trans{\X}$ is shorthand for
$\trans{\X,\X}$, and $\I_{\lin{\X}}\in\trans{\X}$ 
denotes the identity super-operator.

A super-operator $\Phi\in\trans{\X,\Y}$ is
{\it positive} if $\Phi(P)\in\pos{\Y}$ for every $P\in\pos{\X}$, and
is {\it completely positive} if $\Phi\otimes\I_{\lin{\Z}}$ is positive
for every complex Euclidean space $\Z$.
Super-operators that are both completely positive and trace-preserving
will be called {\it admissible} super-operators.
Such super-operators represents valid {\it quantum operations} from
a system with associated space $\X$ to one with associated space $\Y$.

With respect to the standard basis of $\X$, the 
{\it Choi-Jamio{\l}kowski representation} of a super-operator 
$\Phi\in\trans{\X,\Y}$ is defined as
\[
J(\Phi) = \sum_{a,b\in\Sigma} \Phi(\ket{a}\bra{b}) \otimes \ket{a}\bra{b}.
\]
The resulting mapping
$J: \trans{\X,\Y} \rightarrow \lin{\Y\otimes\X}$
is a linear bijection.
It is the case that $\Phi$ is completely positive if and only if
$J(\Phi)$ is positive semidefinite.

Every super-operator $\Phi\in\trans{\X,\Y}$ can be expressed as
\[
\Phi(X) = \sum_{j = 1}^k A_j X B_j^{\ast}
\]
for some choice of an integer $k\geq 1$ and operators
$A_1,\ldots,A_k,B_1,\ldots,B_k\in\lin{\X,\Y}$.
This expression is called a {\it Kraus representation} of $\Phi$ and the
operators $A_1,\ldots,A_k$ and $B_1,\ldots,B_k$ are referred to as
{\it Kraus operators}.
The minimal value of $k$ for which such an expression exists is
$k = \op{rank}(J(\Phi))$.
In case $\Phi$ is completely positive one may take
$A_j = B_j$ for all $j = 1,\ldots,k$.

Finally, every super-operator $\Phi\in\trans{\X,\Y}$ can be expressed as
\[
\Phi(X) = \tr_{\Z} \( A X B^{\ast}\)
\]
for some choice of a complex Euclidean space $\Z$ and operators
$A,B\in\lin{\X,\Y\otimes\Z}$.
In particular, such a representation exists provided 
that $\op{dim}(\Z)\geq \op{rank}(J(\Phi))$.
When $\Phi$ is completely positive one may take $A = B$, and such an
expression is called a {\it Stinespring representation} of $\Phi$.

\subsection{Distinguishability of quantum operations}

The trace distance between quantum states directly relates to their
distinguishability.
This relation can be simply expressed by referring to the following
abstract problem.
\begin{problem}[\it Distinguishing quantum states]
Quantum states $\rho_0,\rho_1\in\density{\X}$ are fixed, and
a bit $a\in\{0,1\}$ is chosen uniformly at random.
The goal is to guess the value of $a$ with probability as large as
possible by means of a measurement of a single copy of $\rho_a$.
\end{problem}

The optimal probability to correctly guess $a$ is precisely
\[
\frac{1}{2} + \frac{1}{4}\tnorm{\rho_0 - \rho_1}.
\]
Indeed, any measurement performed on $\rho_0$ and $\rho_1$ will result
in probability mass functions $p_0$ and $p_1$ for which 
$\norm{p_0-p_1}_1 \leq \tnorm{\rho_0 - \rho_1}$, and moreover equality
is achieved by a two-outcome (projective) measurement. 

As briefly discussed in the introduction, we may consider a similar
problem for quantum operations rather than states.
\begin{problem}[\it Distinguishing quantum operations]
Quantum operations $\Phi_0,\Phi_1\in\trans{\X,\Y}$ are
fixed, and a bit $a\in\{0,1\}$ is chosen uniformly at random.
The goal is to guess the value of $a$ with probability as large as
possible by means of a process involving just a single evaluation of
the operation $\Phi_a$.
\end{problem}

The super-operator norm that is most relevant to this problem is
sometimes known as the {\it diamond norm}.
It is defined as follows.

\begin{definition}
Let $\X$ and $\Y$ be complex Euclidean spaces.
For every $\Phi\in\trans{\X,\Y}$, we define the 
{\it super-operator trace norm} of $\Phi$ as
\[
\tnorm{\Phi} \defeq \max\left\{
\tnorm{\Phi(X)} \,:\, X\in\lin{\X},\,\;\tnorm{X}\leq 1\right\},
\]
and we define the {\it diamond norm} of $\Phi$ as 
\[
\dnorm{\Phi} \defeq \tnorm{\Phi\otimes \I_{\lin{\X}}}.
\]
\end{definition}

Let us note that for a given super-operator $\Phi\in\trans{\X,\Y}$,
we have
\[
\tnorm{\Phi} 
= \max\left\{\tnorm{\Phi(u v^{\ast})}\,:\,u,v\in\sphere{\X}\right\}
\]
and therefore
\[
\dnorm{\Phi} 
= \max\left\{\tnorm{(\Phi\otimes \I_{\lin{\X}})(u v^{\ast})}\,:\,
u,v\in\sphere{\X\otimes\X}\right\}.
\]
It holds that
\[
\dnorm{\Phi} = \tnorm{\Phi\otimes \I_{\lin{\Z}}}
\]
for any choice of $\Z$ whose dimension is at least that of $\X$.

The diamond norm, first used in the setting of quantum information by
Kitaev \cite{Kitaev97}, has precisely the same relationship to the
problem of distinguishing quantum operations as the trace norm has to
distinguishing quantum states.
Specifically, the quantity $\dnorm{\Phi_0 - \Phi_1}$ represents the
maximal $\ell_1$-distance between two probability distributions
resulting from {\it interactive measurements} of the operations
$\Phi_0$ and $\Phi_1$, where an interactive measurement refers to the
process of preparing a state, evaluating a quantum operation on part
of that state, and measuring the result.
In particular, the optimal probability to correctly guess the value of
the bit $a$ in the problem above is
\[
\frac{1}{2} + \frac{1}{4}\dnorm{\Phi_0 - \Phi_1}.
\]
Roughly speaking, the inclusion of the tensor factor $\I_{\lin{\X}}$
in the definition of the diamond norm accounts for the use of an
auxiliary space in a process that attempts to distinguish between
super-operators.
It should be appreciated, however, that the diamond norm happens to
be very robust and possesses nice properties that also contribute
to its use for this application.

The diamond norm is closely related to the 
{\it norm of complete boundedness}, which plays an important role in
operator theory \cite{Paulsen02} and is sometimes referenced in
quantum information theory.
Specifically, it holds that
$\dnorm{\Phi} = \cbnorm{\Phi^{\ast}}$
for any super-operator $\Phi\in\trans{\X,\Y}$, where
$\Phi^{\ast}\in\trans{\Y,\X}$ denotes the adjoint super-operator to
$\Phi$.
It must be kept in mind, however, that the norm of complete
boundedness (as it is most commonly defined) gives an appropriate way
to measure distance between quantum operations in the so-called 
{\it Heisenberg picture} formulation of quantum information and not in
the more common {\it Schr{\"o}dinger picture} formulation; for it is
the quantity $\dnorm{\Phi_0-\Phi_1}=\cbnorm{\Phi_0^{\ast}-\Phi_1^{\ast}}$
and not $\cbnorm{\Phi_0-\Phi_1}$ that directly relates to the
distinguishability of $\Phi_0$ and $\Phi_1$ in the sense discussed
above.

\section{An illustrative example} \label{sec:example}

A simple example was presented in the introduction illustrating the
use of entanglement to distinguish admissible super-operators.
In that example, the use of an entangled input allows perfect
distinguishability of two quantum operations that can be distinguished
correctly with probability at most 5/6 without the use of entangled
inputs.
In this section we present a class of examples that show a more
striking difference between strategies that entangle inputs with an
auxiliary system and those that do not.
A similar example appears in \cite{KretschmannSW06}.

Let $\X$ be a complex Euclidean space and let $n = \op{dim}(\X)$.
Define admissible super-operators $\Phi_0,\Phi_1\in\trans{\X}$ as follows:
\begin{align*}
\Phi_0(X) & = \frac{1}{n+1}\( (\tr X) \I_{\X} + X^{\t}\),\\[2mm]
\Phi_1(X) & = \frac{1}{n-1}\( (\tr X) \I_{\X} - X^{\t}\).
\end{align*}
Here, $X^{\t}$ denotes transposition with respect to the standard
basis of $\X$.
It is clear from the definitions that both $\Phi_0$ and $\Phi_1$ are
trace-preserving, while complete positivity follows from a calculation
of the Choi-Jamio{\l}kowski representations of these super-operators:
\[
J(\Phi_0) = \frac{2}{n+1} \Pi_{\X\ovee\X} \quad\text{and}\quad
J(\Phi_1) = \frac{2}{n-1} \Pi_{\X\owedge\X},
\]
where $\X\ovee\X$ and $\X\owedge\X$ are the symmetric and
antisymmetric subspaces of $\X\otimes\X$, respectively.

These two operations can be distinguished perfectly, provided that a
sufficiently large auxiliary quantum system is used.
To see this, consider these operations applied to half of the maximally
entangled state
\[
\xi = \frac{1}{n}\sum_{a,b\in\Sigma}\ket{a}\bra{b} \otimes
\ket{a}\bra{b}
\in\density{\X\otimes\X}.
\]
We have
\[
(\Phi_0\otimes \I_{\lin{\X}})(\xi) 
= \frac{2}{n(n+1)}\Pi_{\X\ovee\X} \quad\text{and}\quad
(\Phi_1\otimes \I_{\lin{\X}})(\xi) 
= \frac{2}{n(n-1)}\Pi_{\X\owedge\X}.
\]
As $\X\ovee\X$ and $\X\owedge\X$ are orthogonal, it holds that
\[
\tnorm{(\Phi_0\otimes \I_{\lin{\X}})(\xi)
-(\Phi_1\otimes \I_{\lin{\X}})(\xi)} = 2.
\]
This implies that the density operators $(\Phi_0\otimes
\I_{\lin{\X}})(\xi)$ and $(\Phi_1\otimes \I_{\lin{\X}})(\xi)$,
and therefore the super-operators $\Phi_0$ and $\Phi_1$, can be
distinguished without error.

Now suppose $\W_k$ represents an auxiliary space of dimension
$k$, where $1\leq k \leq n$.
It is clear by convexity that the quantity
\[
\tnorm{(\Phi_0\otimes \I_{\lin{\W_k}})(\rho)
-(\Phi_1\otimes \I_{\lin{\W_k}})(\rho)}
\]
is maximized for $\rho = u u^{\ast}$, where $u\in\X\otimes\W_k$ is a unit
vector.
Fix such a vector $u$, and write
\[
u = \sum_{j = 1}^k \sqrt{p_j} x_j \otimes w_j
\]
for $\{x_1,\ldots,x_k\}\subset\X$ and $\{w_1,\ldots,w_k\}\subset\W_k$
orthonormal sets and $p_1,\ldots,p_k\geq 0$.
Noting that
\begin{align*}
(\Phi_0\otimes \I_{\lin{\W_k}})(u u^{\ast})
-(\Phi_1\otimes \I_{\lin{\W_k}})(u u^{\ast})
\hspace{-6cm} \\
& =
\frac{2}{n^2 - 1}
\sum_{j = 1}^k
p_j \left(n \overline{x_j} x_j^{\t} - \I_{\X}\right) \otimes
w_j w_j^{\ast}
+ \frac{2n}{n^2 - 1}
\sum_{i\not=j}
\sqrt{p_i p_j}\:
\overline{x_j} x_i^{\t} \otimes w_i w_j^{\ast}
\end{align*}
provides a simple upper bound:
\[
\tnorm{
(\Phi_0\otimes \I_{\lin{\W_k}})(u u^{\ast})
-(\Phi_1\otimes \I_{\lin{\W_k}})(u u^{\ast})}
\leq \frac{4}{n+1} + \frac{2n}{n^2 - 1}(k-1).
\]

This inequality is obviously not tight for some values of $k$; but it
nevertheless shows that any significant decrease in the size of the
auxiliary space results in a significant error in distinguishing these
super-operators.
In particular, by taking $k=1$ we see that the quantum operations
$\Phi_0$ and $\Phi_1$ act nearly identically on input states that are
not entangled with an auxiliary system.

\section{The main result} \label{sec:main}

This section contains a proof of the bound discussed in the
introduction.
A formal statement of this result is given in the following theorem.

\begin{theorem} \label{theorem:main}
Let $\X$ and $\Y$ be complex Euclidean spaces, let
$\Phi_0,\Phi_1\in\trans{\X,\Y}$ be admissible super-operators, and let
$k = \op{rank}(J(\Phi_0 - \Phi_1))$.
Then for any complex Euclidean space $\W$ with 
$\op{dim}(\W)\geq 2k$ there exists a unit vector
$u\in\X\otimes\W$ such that
\[
\tnorm{
(\Phi_0 \otimes \I_{\lin{\W}})(u u^{\ast})
-
(\Phi_1 \otimes \I_{\lin{\W}})(u u^{\ast})
}
=
\dnorm{\Phi_0 - \Phi_1}.
\]
\end{theorem}

\noindent
Before proceeding to the proof of this theorem, let us briefly discuss
its interpretation in terms of the super-operator distinguishability
problem. 

We suppose that we are given admissible super-operators $\Phi_0$ and
$\Phi_1$ mapping $\lin{\X}$ to $\lin{\Y}$, and that these
super-operators are to be distinguished in the sense of the abstract
problem discussed previously.
Let $k = \op{rank}(J(\Phi_0 - \Phi_1))$, which is at most the sum of
the number of Kraus operators needed to express $\Phi_0$ and $\Phi_1$.

We know that the optimal probability to distinguish the
super-operators, by which we mean the optimal probability to
correctly identify $\Phi_a$ for $a\in\{0,1\}$ chosen uniformly,~is
\[
\frac{1}{2} + \frac{1}{4}\dnorm{\Phi_0 - \Phi_1}.
\]
The theorem implies it is possible to achieve this probability of
success by preparing some pure state $u \in \X\otimes\W$ for
$\W$ corresponding to an auxiliary system of dimension at most
$2k$, applying $\Phi_a$ to this state, and measuring the result.
This is because an optimally chosen measurement correctly
distinguishes between the states
$(\Phi_0 \otimes \I_{\lin{\W}})(u u^{\ast})$
and $(\Phi_1 \otimes \I_{\lin{\W}})(u u^{\ast})$
with probability
\[
\frac{1}{2} + \frac{1}{4}\tnorm{
(\Phi_0 \otimes \I_{\lin{\W}})(u u^{\ast})
-(\Phi_1 \otimes \I_{\lin{\W}})(u u^{\ast})}
=
\frac{1}{2} + \frac{1}{4}\dnorm{\Phi_0 -\Phi_1}.
\]

The proof of Theorem~\ref{theorem:main} is split into three subsections.
The first subsection establishes a fact about the rank of an
input density operator to a positive super-operator required to yield
a given output.
The second subsection relates the super-operator trace norm and
diamond norm to the maximum output fidelity of completely positive
super-operators.
Finally, the third subsection combines these facts to prove the main
theorem.

\subsection{A theorem on the minimum rank of a preimage}

Let $\Phi\in\trans{\X,\Y}$ be a positive super-operator.
Define
\[
\op{Out}(\Phi) \defeq \left\{ \Phi(\rho)\,:\,\rho\in\density{\X}
\right\}
\]
to be the set of all outputs of $\Phi$ ranging over all density
operator inputs, and for a given operator $P\in\op{Out}(\Phi)$ let us
consider the set
\begin{equation} \label{eq:preimage}
\{\rho\in\density{\X}\,:\,\Phi(\rho) = P\}.
\end{equation}
In this section we prove that this set must include at least one
density operator $\rho$ that satisfies $\op{rank}(\rho) \leq
\op{rank}(P)$, provided that $P\not=0$.
(We really only need this fact for completely positive $\Phi$, but
the proof goes through for all positive $\Phi$.)

The basic idea of the proof is as follows.
We observe that the above set \eqref{eq:preimage} is a nonempty,
compact, and convex, and therefore has at least one extreme point.
Assuming that $P$ is nonzero, it may be argued that any such extreme
point must have rank at most that of $P$.
The proof below is based on the proof of Proposition 13.1 in Chapter
II of Barvinok \cite{Barvinok02}, with some minor refinements possible given
the particular assumptions at hand.

\begin{theorem} \label{theorem:input-rank}
  Let $\X$ and $\Y$ be complex Euclidean spaces and let 
  $\Phi\in\trans{\X,\Y}$ be a positive super-operator.
  Then for every choice of $P\in\op{Out}(\Phi)$ with $P\not=0$ there
  exists a density operator $\rho\in\density{\X}$ such that
  \begin{enumerate}
  \item $\Phi(\rho) = P$, and
  \item $\op{rank}(\rho) \leq \op{rank}(P)$.
  \end{enumerate}
\end{theorem}

\begin{proof}
  Let $n = \op{dim}(\X)$, $m = \op{dim}(\Y)$, and $k = \op{rank}(P)$.
  Using a spectral decomposition of $P$ we may write
  \[
  P = \sum_{i = 1}^k y_i y_i^{\ast}
  \]
  for some orthogonal collection $\{y_1,\ldots,y_k\}\subset\Y$.
  Define $\U = \op{span}\{y_1,\ldots,y_k\}$.

  Next, viewing spaces of Hermitian operators as real vector spaces,
  we define a real linear mapping
  \[
  \Psi : \herm{\X} \rightarrow \herm{\U} \oplus \real
  \]
  as follows.
  For each $X\in \herm{\X}$ we define $\Psi(X) = (Y,\lambda)$, for
  \begin{align*}
    Y & = \Pi_{\U} \Phi(X) \Pi_{\U},\\
    \lambda & = \tr\left[ \(\I_{\X} - \Pi_{\U}\) \Phi(X)\right].
  \end{align*}
  Given that $\herm{\U}\oplus\real$ is a $(k^2 + 1)$-dimensional real
  vector space, it holds that
  \[
  \ker(\Psi) = \left\{X\in\herm{\X}\,:\,\Psi(X) = (0,0)\right\}
  \]
  is a subspace of $\herm{\X}$ having dimension at least $n^2 - \(k^2 + 1\)$.

  For every choice of $\rho\in\density{\X}$ it holds that
  $\Phi(\rho) = P$ if and only if $\Psi(\rho) = (P,0)$, and therefore
  \[
  \left\{\rho\in\density{\X}\,:\,\Phi(\rho) = P\right\}
  = \left\{\rho\in\density{\X}\,:\,\Psi(\rho) = (P,0)\right\}.
  \]
  This set is non-empty, compact, and convex, and we may therefore
  choose an extreme point $\rho$ from this set.
  To complete the proof, it suffices to prove that
  $r = \op{rank}(\rho)\leq k$.
  
  Using a spectral decomposition of $\rho$ we may write
  \[
  \rho = \sum_{i = 1}^r p_i x_i x_i^{\ast}
  \]
  for $p_1,\ldots,p_r > 0$ and $\{x_1,\ldots,x_r\}$ orthogonal unit
  vectors in $\X$.
  Let $\V = \op{span}\{x_1,\ldots,x_r\}$ and let
  $\A\subseteq\herm{\X}$ be the subspace defined as
  \[
  \A = \left\{X\in\herm{\X}\,:\,\op{im}(X)\subseteq\V,\,\tr(X) = 0\right\}.
  \]
  Equivalently, $\A$ is the subspace containing all traceless
  Hermitian operators of the form
  \[
  X = \sum_{1\leq i,j\leq r} \alpha_{i,j} x_i x_j^{\ast}.
  \]
  Observe that $\op{dim}(\A) = r^2 - 1$ (again, as a real vector space).

  Consider the intersection of the subspaces $\ker(\Psi)$ and $\A$,
  and suppose $X\in\ker(\Psi) \cap \A$ is any element of this
  intersection. 
  Our goal will be to prove that $X=0$, and therefore that the
  intersection $\ker(\Psi) \cap \A$ is trivial.
  To this end, assume toward contradiction that $X\not=0$.
  As $X$ is Hermitian and $\op{im}(X) \subseteq\V$, we have that
  \[
  \pm X \leq \norm{X} \Pi_\V.
  \]
  Given that $\delta \Pi_{\V} \leq \rho$ for 
  $\delta = \op{min}(p_1,\ldots,p_r)> 0$, it follows 
  that $\pm \varepsilon X \leq \rho$ for $\varepsilon = \delta/\norm{X}$.
  Because $X$ is traceless, this implies that
  $\rho \pm \varepsilon X\in\density{\X}$.
  Finally, given that $X\in\op{ker}(\Psi)$, we have
  $\Psi(\rho \pm \varepsilon X) = (P,0)$, which is equivalent to
  $\Phi(\rho \pm \varepsilon X) = P$.

  At this point we have proved that
  \[
  \Phi(\rho - \varepsilon X) = \Phi(\rho) = \Phi(\rho + \varepsilon X),
  \]
  and we have that $\rho$, $\rho - \varepsilon X$ and 
  $\rho+\varepsilon X$ are distinct density operators.
  Given that
  \[
  \rho = \frac{1}{2} (\rho - \varepsilon X) + \frac{1}{2} (\rho +
  \varepsilon X)
  \]
  and that $\rho$ was chosen to be an extreme point in the set
  $\left\{\rho\in\density{\X}\,:\,\Phi(\rho) = P\right\}$, we have
  arrived at a contradiction.
  It is therefore established that the subspaces $\ker(\Psi)$ and
  $\A$ have a trivial intersection.

  Now, given that $\op{ker}(\Psi)$ and $\A$ are subspaces of
  $\herm{\X}$ with
  \begin{align*}
    \op{dim}(\op{ker}(\Psi)) & \geq n^2 - \(k^2 + 1\),\\
    \op{dim}(\A) & = r^2 - 1,\\
    \op{dim}(\op{ker}(\Psi)\cap\A) & = 0,
  \end{align*}
  we have 
  $n^2 - \(k^2 + 1\) + \left(r^2 - 1\right) \leq n^2$,
  and therefore $r^2 \leq k^2 + 2$.
  As $r$ and $k$ are positive integers, we conclude that $r \leq k$,
  which completes the proof.
\end{proof}

\begin{remark}
Note that the assumption $P\not=0$ is necessary because a density
operator cannot have zero rank.
It of course follows easily from the positivity of $\Phi$ that if
$\Phi(\rho) = 0$ for some density operator $\rho$, then this is so for
some $\rho$ having rank 1.
This fact also happens to be revealed by the above proof, which really
only uses the assumption that $P\not=0$ at the very end.
In particular, if $k = 0$, the inequality $r^2 \leq k^2 + 2$ only implies
that $r \leq 1$.
\end{remark}

\subsection{Distinguishability and maximum output fidelity}

We now relate the super-operator trace norm and diamond norm to the
fidelity of outputs of completely positive super-operators, maximized
over various sets.
Let us begin with two definitions.

\begin{definition}
For every complex Euclidean space $\X$ and integer $k\geq 1$, define 
\[
\opset{D}{k}{\X} \defeq \left\{
\rho\in\density{\X}\,:\,\op{rank}(\rho) \leq k\right\}.
\]
\end{definition}

\begin{definition}
Suppose $\X$ and $\Y$ are complex Euclidean spaces and
$\Psi_1,\Psi_2\in\trans{\X,\Y}$ are completely positive
super-operators.
For each $k\geq 1$ define
\[
\fid^{(k)}_{\mathrm{max}}(\Psi_1,\Psi_2) \defeq \max\left\{
\fid(\Psi_1(\rho_1),\Psi_2(\rho_2))\,:\,
\rho_1,\rho_2\in\opset{D}{k}{\X}
\right\}.
\]
We also write $\fid_{\mathrm{max}}(\Psi_1,\Psi_2)
=\fid^{(n)}_{\mathrm{max}}(\Psi_1,\Psi_2)$ for $n = \op{dim}(\X)$,
which allows for a maximization over all density operators $\rho_1$
and $\rho_2$ in the above equation.
\end{definition}

\noindent
We will also require the following lemma, proved in \cite{RosgenW05}.
A short proof is included for completeness.

\begin{lemma}\label{lemma:fidelity-tnorm}
Let $\X$ and $\Y$ be complex Euclidean spaces and let
 $P,Q\in\pos{\X}$.
Assume that $u,v\in\X\otimes\Y$ satisfy $\tr_{\Y}(u u^{\ast}) = P$ and
$\tr_{\Y}(v v^{\ast}) = Q$.
Then $\fid(P,Q) = \tnorm{\tr_{\X}(u v^{\ast})}$.
\end{lemma}

\begin{proof}
For any choice of $Y\in\lin{\Y}$ we have
\[
\tnorm{Y} = \max_U \abs{\tr(UY)}
\]
where the maximization is over all unitary operators $U\in\lin{\Y}$,
and therefore
\[
\tnorm{\tr_{\X}(u v^{\ast})} = 
\max_U \abs{\tr\left(U \tr_{\X}(u v^{\ast})\right)}
= \max_U \abs{v^{\ast} \(\I_{\X}\otimes U\) u}.
\]
As $U$ ranges over all possible unitary operators on $\Y$, the vector
$(\I_\X\otimes U)u$ ranges over all purifications of $P$ in
$\X\otimes\Y$.
The above quantity is therefore equal to $\fid(P,Q)$ by Uhlmann's
Theorem (q.v.~Theorem 9.4 in \cite{NielsenC00}).
\end{proof}

Now, the relation between distinguishability and maximum output fidelity
that will established is given by the following theorem
(cf.~Corollary 2.2 of \cite{Timoney07}).

\begin{theorem}  \label{theorem:max-fidelity-k}
  Let $\X$, $\Y$, and $\Z$ be complex Euclidean spaces, 
  let $\Phi\in\trans{\X,\Y}$ be an arbitrary super-operator, and
  suppose that $A,B\in\lin{\X,\Y\otimes\Z}$ satisfy
  $\Phi(X) = \tr_{\Z} (A X B^{\ast})$ for all $X\in\lin{\X}$.
  Define completely positive super-operators
  $\Psi_A,\Psi_B\in\trans{\X,\Z}$ as
  \begin{align*}
    \Psi_A(X) & = \tr_{\Y}\( A X A^{\ast}\),\\
    \Psi_B(X) & = \tr_{\Y}\( B X B^{\ast}\),
  \end{align*}
  for all $X\in\lin{\X}$.
  Then for all $k\geq 1$ it holds that
  \[
  \fid^{(k)}_{\mathrm{max}}(\Psi_A,\Psi_B) = \tnorm{\Phi \otimes
    \I_{\lin{\W_k}}},
  \]
  where $\W_k$ is any complex Euclidean space with dimension $k$.
\end{theorem}

\begin{remark}
Note that it is the space $\Y$ that is traced-out in the definition of
$\Psi_A$ and $\Psi_B$, rather than the space $\Z$.
\end{remark}

\begin{proof}
  Let us fix $k\geq 1$ and let $\W_k$ be a complex Euclidean space of
  dimension $k$.
  For any choice of $u,v\in\X\otimes\W_k$ we have
  \begin{align*}
    \tnorm{(\Phi\otimes \I_{\lin{\W_k}})(u v^{\ast})} \hspace{-2.5cm}\\
    & =
    \tnorm{\tr_{\Z}\left[
	(A \otimes \I_{\W_k})uv^{\ast} (B^{\ast} \otimes \I_{\W_k})
	\right]}\\
    & =
    \fid\left(
    \tr_{\Y\otimes\W_k}\left(
    (A \otimes \I_{\W_k})u u^{\ast} (A^{\ast} \otimes \I_{\W_k})\right),
    \tr_{\Y\otimes\W_k}\left(
    (B \otimes \I_{\W_k})v v^{\ast} (B^{\ast} \otimes
    \I_{\W_k})\right)\right)\\
    & = \fid\(
    \Psi_A(\tr_{\W_k}(u u^{\ast})),
    \Psi_B(\tr_{\W_k}(v v^{\ast}))\),
  \end{align*}
  where the second equality is by Lemma~\ref{lemma:fidelity-tnorm}.
  Given that $\op{dim}(\W_k) = k$, it holds that
  \[
  \left\{\tr_{\W_k}(u u^{\ast})\,:\,u\in\sphere{\X\otimes\W_k}\right\}
  = \opset{D}{k}{\X}.
  \]
  This implies that
  \begin{align*}
    \tnorm{\Phi\otimes \I_{\lin{\W_k}}}
    & = 
    \max\left\{
    \tnorm{(\Phi\otimes \I_{\lin{\W_k}})(u v^{\ast})}\,:\,
    u,v\in\sphere{\X\otimes\W_k}\right\}\\
    & =
    \max\left\{
    \fid\( \Psi_A(\rho_A),\Psi_B(\rho_B)\)\,:\,
    \rho_A,\rho_B\in\opset{D}{k}{\X}\right\}\\
    & = \fid_{\mathrm{max}}^{(k)}(\Psi_A,\Psi_B)
  \end{align*}
  as required.
\end{proof}

The following corollary, which corresponds to the case 
$k = \op{dim}(\X)$ in the previous theorem, is of special interest.
This fact is implicit in \cite{KitaevW00} and appears (as an exercise)
in \cite{KitaevS+02}.

\begin{cor} \label{cor:max-fidelity}
  Suppose that $\Phi\in\trans{\X,\Y}$ and
  $\Psi_A,\Psi_B\in\trans{\X,\Z}$ are as in
  Theorem~\ref{theorem:max-fidelity-k}.
  Then
  \[
  \fid_{\mathrm{max}}(\Psi_A,\Psi_B) = \dnorm{\Phi}.
  \]
\end{cor}

\subsection{Optimal distinguishability with small auxiliary systems}

Now we combine the results of the previous two subsections to bound
the size of the auxiliary space needed to optimally distinguish
quantum operations.
First we prove the following theorem.

\begin{theorem} \label{theorem:trace-diamond}
Let $\Phi\in\trans{\X,\Y}$ be a super-operator, let
$k = \op{rank}(J(\Phi))$, and let $\W_k$ be a complex Euclidean
space having dimension $k$.
Then
\[
\dnorm{\Phi} = \tnorm{\Phi\otimes\I_{\lin{\W_k}}}.
\]
\end{theorem}

\begin{proof}
As $\op{rank}(J(\Phi)) = k$, we may write
$\Phi(X) = \tr_{\W_k}\left(A X B^{\ast}\right)$
for $A,B\in\lin{\X,\Y\otimes\W_k}$.
By Corollary~\ref{cor:max-fidelity},
$\dnorm{\Phi} = \fid_{\mathrm{max}}(\Psi_A,\Psi_B)$
for $\Psi_A,\Psi_B\in\trans{\X,\W_k}$ defined as
\begin{align*}
\Psi_A (X) & = \tr_{\Y} \left( A X A^{\ast}\right),\\
\Psi_B (X) & = \tr_{\Y} \left( B X B^{\ast}\right).
\end{align*}
Let $\rho_A,\rho_B\in\density{\X}$ be density operators that achieve
this maximum fidelity:
\[
\fid_{\mathrm{max}}(\Psi_A,\Psi_B) = \fid(\Psi_A(\rho_A),\Psi_B(\rho_B)).
\]

The operators $\Psi_A(\rho_A)$ and $\Psi_B(\rho_B)$ are contained in
$\pos{\W_k}$, and therefore have rank at most~$k$.
By Theorem~\ref{theorem:input-rank} there must therefore exist density
operators $\xi_A,\xi_B\in\density{\X}$ having rank at most~$k$ such
that $\Psi_A(\xi_A) = \Psi_A(\rho_A)$ and $\Psi_B(\xi_B) = \Psi_B(\rho_B)$.
Thus
\[
\fid_{\mathrm{max}}^{(k)}(\Psi_A,\Psi_B) 
\geq \fid(\Psi_A(\xi_A),\Psi_B(\xi_B))
= \fid(\Psi_A(\rho_A),\Psi_B(\rho_B))
= \fid_{\mathrm{max}}(\Psi_A,\Psi_B).
\]
The reverse inequality obviously holds, and so by
Theorem~\ref{theorem:max-fidelity-k} and Corollary~\ref{cor:max-fidelity}
we have
\[
\dnorm{\Phi} 
= \fid_{\mathrm{max}}(\Psi_A,\Psi_B)
= \fid_{\mathrm{max}}^{(k)}(\Psi_A,\Psi_B) 
= \tnorm{\Phi\otimes \I_{\lin{\W_k}}}
\]
as required.
\end{proof}

Before completing the proof Theorem~\ref{theorem:main}, we need one
more lemma.
It is similar to Lemma~2.4 in \cite{RosgenW05}, but is slightly more
general.
We need this lemma because the value of the super-operator trace norm
is not always achieved by a density operator input, even when the
super-operator is the difference between admissible super-operators
\cite{Watrous05}.

\begin{lemma} \label{lemma:trace-norm-hermitian}
Let $\Phi = \Phi_0 - \Phi_1$ for completely positive super-operators
$\Phi_0,\Phi_1\in\trans{\X,\Y}$, and let $\Q$ be a complex Euclidean
space with dimension 2.
Then there exists a unit vector $u \in \X\otimes \Q$ such that
\[
\tnorm{(\Phi\otimes\I_{\lin{\Q}})(u u^{\ast})}
\geq \tnorm{\Phi}.
\]
\end{lemma}

\begin{proof}
Let $X\in\lin{\X}$ be an operator with $\tnorm{X} = 1$ that
satisfies $\tnorm{\Phi} = \tnorm{\Phi(X)}$, and define
\[
Y = \frac{1}{2}X\otimes \ket{0}\bra{1} + \frac{1}{2}X^{\ast}\otimes
\ket{1}\bra{0}\in\herm{\X\otimes\Q}.
\]
Here, we assume the standard basis of $\Q$ is $\{\ket{0},\ket{1}\}$.
Then $\tnorm{Y} = \tnorm{X} = 1$ and
\begin{align*}
  \tnorm{(\Phi \otimes \I_{\lin{\Q}})(Y)}
  & = \frac{1}{2} \tnorm{ \Phi(X) \otimes \ket{0}\bra{1} + 
    \Phi(X^{\ast}) \otimes \ket{1}\bra{0}}\\
  & = \frac{1}{2} \tnorm{\Phi(X) \otimes \ket{0}\bra{1} + 
    \Phi(X)^{\ast} \otimes \ket{1}\bra{0}}\\
  & = \tnorm{\Phi(X)}\\
  & = \tnorm{\Phi}.
\end{align*}
The second equality follows from the condition that
$\Phi = \Phi_0 - \Phi_1$ for $\Phi_0$ and $\Phi_1$
completely positive, which is equivalent to
$\Phi(X^{\ast}) = \Phi(X)^{\ast}$ for all $X\in\lin{\X}$.

Now, because $Y$ is Hermitian, we may consider a spectral decomposition
\[
Y = \sum_i \lambda_i u_i u_i^{\ast}.
\]
By the triangle inequality, it holds that
\[
\tnorm{\Phi} = \tnorm{(\Phi \otimes \I_{\lin{\Q}})(Y)}
\leq \sum_i \abs{ \lambda_i } \tnorm{(\Phi \otimes \I_{\lin{\Q}})
(u_i u_i^{\ast})}.
\]
As $\tnorm{Y} = 1$, we have $\sum_i |\lambda_i| = 1$, and thus
\[
\tnorm{(\Phi \otimes \I_{\lin{\Q}})(u_i u_i^{\ast})}
\geq \tnorm{\Phi}
\]
for some index $i$.
Setting $u = u_i$ for any such choice of $i$ completes the proof.
\end{proof}

Finally we have all of the facts that we require to prove
Theorem~\ref{theorem:main}.
The proof follows.

\begin{proof}[Proof of Theorem~\ref{theorem:main}]
By Theorem~\ref{theorem:trace-diamond} it follows that
\[
\dnorm{\Phi_0 - \Phi_1}
= \tnorm{\Phi_0\otimes \I_{\lin{\V}} - \Phi_1\otimes \I_{\lin{\V}}}
\]
for any complex Euclidean space $\V$ having dimension at least $k$.
By Lemma~\ref{lemma:trace-norm-hermitian} there exists a unit vector
$u \in \X\otimes\V\otimes\Q$ such that
\[
\tnorm{
(\Phi_0 \otimes \I_{\lin{\V\otimes\Q}})(u u^{\ast})
-
(\Phi_1 \otimes \I_{\lin{\V\otimes\Q}})(u u^{\ast})}
\geq \dnorm{\Phi_0 - \Phi_1},
\]
where $\Q$ is any space with dimension 2.
The reverse inequality holds due to a general property of the diamond
norm.
Taking $\W = \V\otimes\Q$ establishes the theorem.
\end{proof}

\subsection*{Acknowledgments}

I thank David Kribs and Vern Paulsen for bringing the work of Timoney
to my attention, and Gus~Gutoski for helpful comments.
This work was supported by Canada's NSERC and the Canadian Institute for
Advanced Research (CIFAR).


\bibliographystyle{alpha}


\end{document}